\documentclass[twocolumn,showpacs,preprintnumbers,amsmath,amssymb]{revtex4}

\usepackage{graphicx}
\begin{document}

\title{Theory of vortex states in magnetic nanodisks with
 induced Dzyaloshinskii-Moriya interactions}


\author{A.B. Butenko$^{1,2}$, A.A. Leonov$^{1,2}$, 
A.N.\ Bogdanov$^{1}$,  U.K. R\"o\ss ler$^{1}$}

\address{$^1$IFW Dresden, Postfach 270116, D-01171 Dresden, Germany}

\address{$^2$Donetsk Institute for Physics and Technology, 
R. Luxemburg 72, 83114 Donetsk, Ukraine}
\date{\today}

\begin{abstract}
{
Vortex states in magnetic
nanodisks are essentially affected by surface/interface
induced Dzyaloshinskii-Moriya interactions.
Within a micromagnetic approach we calculate 
the equilibrium sizes and shape of the vortices
as functions of magnetic field, the material
and geometrical parameters of nanodisks.
It was found that  the Dzyaloshinskii-Moriya
coupling can considerably increase sizes of vortices 
with "right" chirality and suppress vortices 
with opposite chirality. 
This allows to form a bistable system of homochiral vortices
as a basic element for storage applications.
%

}
\end{abstract}


         
\maketitle


\section{Introduction}

Physics of magnetism at nano- and submicrometer 
scales is an object of broad and intensive 
scientific investigations  stimulated by
various possible applications including magnetic random 
access memory, high-density magnetic recording media, 
and magnetic sensors.\cite{Prinz98}
Nanoscale magnetic 
dots with vortex states
are considered as promising
components for such spintronic
devices.\cite{Bussmann99,Zhu00,Bohlens08}
Numerous experimental
observations show that such vortices 
consist of a narrow core with a
perpendicular magnetization surrounded
by an extended area with in-plane magnetization
curling around the center (Fig. \ref{fig1}).
\cite{Shinjo00,Wachowiak02,Raabe00,
Schneider00,Schneider01}
It has  been proposed to use both 
the up and down polarity, 
i.e. the perpendicular magnetization
of a vortex, or the rotation sense 
of the curling in-plane magnetization
as switchable bit elements in memory devices.\cite{Bussmann99,Zhu00,Bohlens08}

The vortex state in thin film elements
results from the necessity to reduce 
the demagnetization energy in competition
with the exchange coupling.\cite{Feldtkeller65,Hubert}
In circular disks the axisymmetric 
ground state arises for diameters 
in the range of a few nanometer up to many 10~nm depending
on magnetic material. For very small diameters, a single domain
state occurs, very large film elements form multidomain states.
Within the usual micromagnetic calculations 
the shape and size of the vortices are 
determined by the competition between exchange 
and stray field energy.\cite{Guslienko08}
In particular, the vortices with different chirality
are degenerate: the four possible vortex ground-states 
differentiated by their handedness and polarity all 
have the same energy within these standard 
micromagnetic models. 

The broken inversion symmetry at the 
surface of magnetic films, however, induces
chiral magnetic couplings in the form
of Dzyaloshinskii-Moriya (DM) exchange as a 
result of relativistic spin-orbit interactions.\cite{Fert90,Crepieux98,PRL01}
The chiral DM interactions destabilize
collinear magnetic states and are able to 
create a large variety of helical and Skyrmionic
spin textures.\cite{Dz64,ANB89,JMMM94,PRL01,Nature06}
The mechanism and phenomenological models of the
surface-induced DM couplings,
along with possible observable effects in magnetic films, 
have been discussed earlier.\cite{Crepieux98,PRL01}
These theories now are supported by modern
quantitative ab initio calculations for
magnetic nanostructures.\cite{Udvardi03,Heide06,Antal08,Mankovsky09}
Recent experiments \cite{Bode07,Ferriani08} provide clear evidence 
for these surface-induced DM interactions, 
as they display long-period modulated non-collinear
magnetic states, which can be identified as 
chiral {\em Dzyaloshinskii spirals}.\cite{Dz64}
Chiral effects observed for magnetization
processes in vortex states of magnetic
nanodisks \cite{Curcic08}
may also belong to this class
of phenomena. 

In this work, we describe the chiral symmetry breaking in 
the vortex ground states of circular thin film
elements within a basic micromagnetic approach.
As the vortex states are chiral themselves, 
the effect of the chiral DM is less obvious. 
However, in the presence of DM interactions
the chiral degeneracy of the left- and right-handed
vortices is lifted.
The simplicity of the circular vortex structure makes them 
amenable to detailed theoretical investigations.
Here, we calculate the differences between 
the core shapes and sizes of left- and right-handed vortices
in the presence of DM couplings.
These differences of core structure may be observable
in experiments, e.g. as differences in core diameter
or net polarity of vortices, when switching their chirality.
We suggest that such experiments can be used to determine
the magnitude of surface-induced DM couplings 
in ultrathin magnetic films/film elements.

\section{Equations and methods}

The energy density of a uniaxial ferromagnet
with chiral interactions can be written 
in the following form \cite{JMMM94}
\begin{eqnarray}
 w &=& A \sum_{i,j}\left(\frac{\partial m_j}{\partial x_i}\right)^2
 -\mathbf{M}\cdot\mathbf{H}
  -\frac{1}{2} \mathbf{M}\cdot\mathbf{H}_m
 \nonumber \\
& &+K_u (\mathbf{m}\cdot\mathbf{a})^2 +w_D,
\label{density}
\end{eqnarray}
where $\mathbf{m}$ is the unity
vector along the magnetization 
$\mathbf{M}=M_s\mathbf{m}$ and
$M_{s}$ is the saturation magnetization.
The couplings are given by the exchange stiffness $A$.%
The anisotropy axis $\mathbf{a}$ is 
taken perpendicularly to the
disk surface, and $K_u$ is the anisotropy constant 
which must be positive in easy-plane materials.
$\mathbf{H}$ is the external magnetic field,
and $\mathbf{H}_m$ is the stray field.
The Dzyaloshinskii-Moriya energy $w_D$
is described by so-called Lifshitz invariants,\cite{Dz64}
energy terms linear in first spatial
derivativs of the magnetization,
\begin{equation}
\label{lifshitz}
L_{ij}^{(k)} = m_i\frac{\partial m_j}{\partial x_k} 
- m_j \frac{ \partial m_i}{\partial x_k}\,.
\end{equation}
The functional form of this energy is determined
by the symmetry of the surface/interfaces.\cite{Crepieux98}
In this paper we use $w_D$ terms in the following form
\begin{equation}
w_D =  D \left(L_{zx}^{(y)}-L_{zy}^{(x)}\right)\,,
\label{lifshitz2} 
\end{equation}
which gives the allowed Lifshitz invariants of 
the magnetization in symmetries from Laue classes 32, 42, and 62.
This $w_D$ term favours the curling mode of the magnetization,
where the rotation sense is determined by the sign of 
the Dzyaloshinskii constant $D$. 
Depending on crystallographic symmetry, other types of Lifshitz
invariants may occur which may favour differently 
twisted non-collinear magnetization structures. 
For example, Lifshitz invariants $(L_{zx}^{(x)}+L_{zy}^{(y)})$,
possible in Laue classes 3m, 4m, and 6m,
induce a cycloidal rotation of the magnetization vector.\cite{JMMM94}
The various possibilities can be deduced from the corresponding
list of Lifshitz invariants 
for three-dimensional Laue classes in Ref.\cite{ANB89},
where also the structures of the corresponding Dzyaloshinskii spirals
and circular vortex-like Skyrmions have been presented.
For simplicity, we restrict discussion 
here to the form of Eq.~(\ref{lifshitz}).
\begin{figure}
\includegraphics[width=7.5cm]{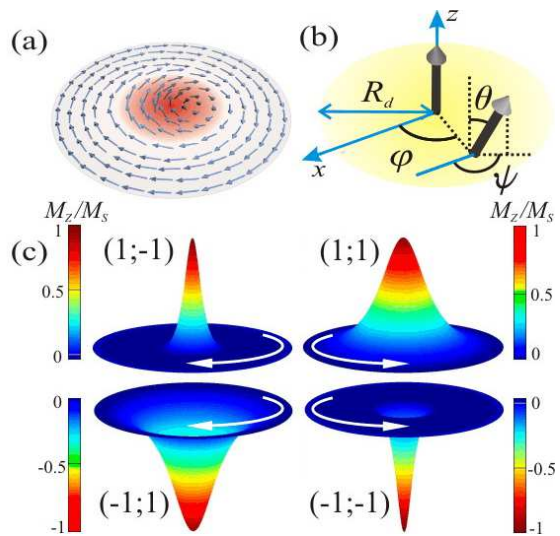}
\caption{ (a) Vortex states in a circular magnetic nanodisk
of radius $R_d$ with  axisymmetric magnetization structure.
(b) Geometry and definition of variables of the problem.
(c) The four possible vortex states 
are characterized by the indices
(\textit{polarity} $p = \pm 1$; \textit chirality $c=\pm 1$)
Dzyaloshinskii-Moriya couplings
with chirality $\tilde{c} =1$ ($D > 0$)
widen vortices with the same
chirality, $c =1$,
and  squeeze vortices with
opposite chirality, $c =-1$.
\label{fig1}
}
\end{figure}

The equilibrium configurations of $\mathbf{m}$
are derived by solving the equations minimizing
the energy (\ref{density}) together with equations
of magnetostatics.
To describe vortex states in a disk of radius
$R_d$ and with zero or perpendicular 
applied field, we consider axisymmetric distributions of
the magnetization and express the magnetization vector 
$\textbf{m}$ in terms of spherical coordinates
and the spatial variables in cylindrical coordinates: 
$\mathbf{m}=(\sin\theta\cos\psi;\sin\theta\sin\psi;\cos\theta)$,
$\mathbf{r}=(\rho\cos\varphi;\rho\sin\varphi;z)$
(Fig. \ref{fig1}, b). 
The vortices are characterized by the magnetization direction
in the center, \textit{polarity} $p = \pm 1$, and
the chirality of the magnetization structure, \textit{chirality} $c = \pm 1$.
Four different vortex states can be described by
the indices ($p; c$) (Fig. \ref{fig1}, c).
One can introduce the \textit{chirality}
of the Dzyaloshinskii-Moriya coupling
as $ D = |D| \tilde{c}$.
Then the Dzyaloshinskii-Moriya interactions
favours  states with  $c = \tilde{c}$
and suppresses those with opposite
chirality $c = - \tilde{c}$.
In this paper we assume for definiteness
that  the vortices have positive chirality, $c =1$,
and study their properties for $D{{<}\atop{>}}\,0$.
Thus, \textit{positive} DM couplings
favour these vortices, while they are suppressed 
in the opposite case, $D < 0$.

Due to the non-local character of stray-field
interactions the micromagnetic problem Eq.~(\ref{density})
constitutes a set of integro-differential equations.\cite{Hubert}
In order to simplify this problem we consider
the limit of a thin film where the magnetodipole
energy has a local character and reduces to
a ``shape'' anisotropy $K_m = 2\pi M_s^2$.\cite{Gioia97} 
This can be added to the uniaxial anisotropy $K_u$
yielding a redefinition of the anisotropy energy 
in Eq.~(\ref{density}) by an effective anisotropy constant $K$.
We also introduce the characteristic (exchange) 
length $l_e$, the anisotropy field $H_a$, 
and a critical value of the Dzyaloshinskii 
constant $D_0$
\begin{eqnarray}
l_{e}=\sqrt{A/K},
\quad
H_a=2K/M_s, 
\nonumber \\
D_0=\sqrt{AK},
\quad
K = K_u + 2\pi M_s^2 > 0,
\label{units}
\end{eqnarray}
as proper material parameters of the problem.
They establish important relations between
vortex solutions and magnetic states
in laterally infinite magnetic nanolayers.
The anisotropy field determines the equilibrium
magnetization of homogeneously magnetized layers 
in an applied perpendicular field $H$,
\begin{equation}
\cos \theta_h = H/H_a\,.
\label{thetah}
\end{equation}
The constant $D_0$ gives 
a threshold "strength" of the DM coupling 
in comparison with the exchange and anisotropy:
for $|D|/D_0 > 4/\pi  = 1.273$ the magnetization of a layer
transforms into a modulated state.\cite{Dz64,JMMM94}
The exchange length gives a characteristic
radius of the vortex core.
Most experimentally investigated nanodisks 
have radii much larger than the exchange length,  $R_d \gg l_e$.
In this case vortices consist
of a strongly localized core encircled by
a wide ring with a constant 
polar angle $\theta= \theta_h$ (Fig.~\ref{fig1}~(a)).

The  variational problem for functional 
(\ref{density}) has rotationally symmetric 
solutions $\psi=\varphi\pm\pi/2$, $\theta = \theta(\rho)$.
By substituting the solution for $\psi$ into
Eq.~(\ref{density}) and integrating
with respect to $\varphi$ the vortex
energy can be reduced to the following
form $E=2\pi\int^{R_d}_{0} w(\rho)\rho d \rho$,
where
 \begin{eqnarray}
w(\rho)&=& A \left[\left(\frac{d \theta}{d \rho}\right)^{2}+\frac{1}{\rho ^2}\sin^2\theta \right]
+K \cos^2\theta
\\ \nonumber
&-&HM_s\cos \theta-D \left(\frac{d \theta}{d \rho}+\frac{1}{\rho}\cos\theta\sin\theta\right),
\label{density2}
\end{eqnarray}
where the magnetic field $H$ is assumed 
to be perpendicular to the disk plane.

The Euler equation for $\theta(\rho)$
\begin{eqnarray}
\label{eq}
A \left( \frac{d^2 \theta}{d \rho ^2} + \frac{1}{\rho}\frac{d \theta}{d \rho}
 -\frac{1}{\rho ^2}\sin \theta \cos \theta  \right)
 -\frac{D}{\rho} \sin^2 \theta
\nonumber  \\ 
+ K \sin \theta \cos \theta 
- H M_s \sin \theta/2  =0
\end{eqnarray}
with the boundary conditions 
\begin{eqnarray}
\theta(0)=0,\, \quad \left(d \theta / d \rho \right)_{\rho =R_d}= g(\theta, R_d)
\label{Boundary}
\end{eqnarray}
yield the equilibrium vortex profiles.
$g(\theta, R_d)$ describes the anchoring 
effect imposed by
the surface energy at the disk edge. 
For $D = H = g(\theta, R_d)  = 0$ and infinite
radius, Eq. (\ref{eq}) is related
to the differential equation 
introduced by Ginzburg
and Pitaevskii in their
theory of  superfluid vortices
in liquid helium.\cite{Ginzburg58}
Similar equations describe vortex excitations
in different bosonic systems including
Bose-Einstein condensates 
(see e.g. \cite{Feder99}) and
easy-plane magnets.\cite{Gouvea89}
\begin{figure}
\includegraphics[width=7.5cm]{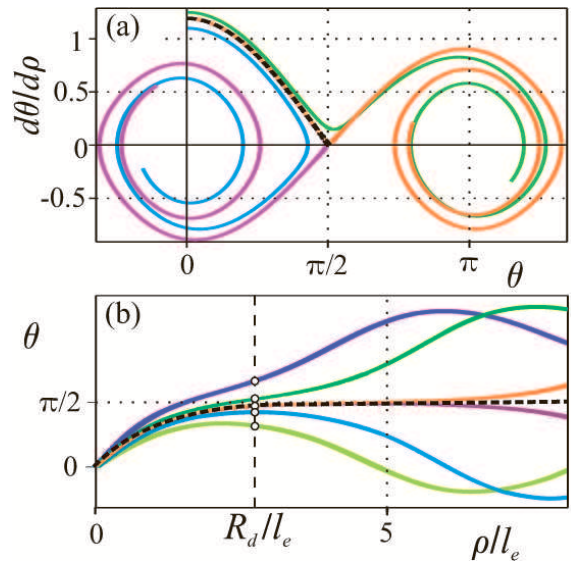}
\caption{ (a) Typical phase space trajectories 
for the solutions of the auxiliary
Cauchy problem (Eqs.~(\ref{eq}), (\ref{Cauchy})).
(b) Corresponding profiles $\theta(\rho)$.
The shown trajectories pertains to the case $H=0$,
$\theta_h=\pi/2$.
Due to the localized character of the vortex states
the equilibrium solutions $\theta(\rho)$ 
of the boundary problem (\ref{eq}), (\ref{Boundary})
are close to the localized profiles given by the (dashed) 
separatrix lines in the phase portrait.
\label{shoot1}
}
\end{figure}

As a first step in solving the boundary value problem 
(Eqs. (\ref{eq}), (\ref{Boundary})) we
consider an  \textit{auxiliary} Cauchy
problem for Eq. (\ref{eq}) with initial conditions
\begin{equation}
\theta(0)=0, \quad  (d\theta/ d \rho)_{\rho = 0} =\alpha
\label{Cauchy}
\end{equation}
for different values of $\alpha>0$.
The trajectories $\theta(\rho; \alpha)$ for 
this initial value problem with $0<\alpha<\infty$ 
define a family of solutions 
parametrized by $\alpha$.
Any solution $\theta(\rho)$ of the boundary problem (\ref{eq}), 
(\ref{Boundary}) is member of this family for a certain 
value of $\alpha$.
A qualitative analysis of possible trajectories 
in the phase space, $d\theta/d\rho\equiv \theta_{\rho}$ vs. $\theta$,
makes it possible to single out the desired solution 
among other trajectories.
Typical trajectories $\theta_{\rho}(\theta)$
for the Cauchy problem are presented in 
the phase space portrait shown in Fig.~\ref{shoot1}.
For arbitrary values of $\alpha$ the lines 
$\theta_{\rho}(\theta)$ usually end by spiraling
around one of the attractors $(\pi n,0)$, 
$n=\pm1,\pm2\dots$.
But for a certain value $\tilde{\alpha}$ the line 
$\theta_{\rho}(\theta)$ ends in the point $(\theta_h,0)$.
Thus, variation of parameter $0 < \alpha  < \infty$ (\ref{Cauchy}) 
allows to  select the solutions of the boundary problem 
(\ref{eq}), (\ref{Boundary}).
The trajectories $\theta(\rho)$
have arrow-like shape.
Their core sizes can be introduced in a
manner commonly used for magnetic domain
walls \cite{Hubert} and Skyrmions
\cite{JMMM94}: as the point $R_0$ where the
tangent at the origin point 
intersects the line $\theta = \theta_h$,
\begin{equation}
R_0 =  \theta_h  \left( d\theta / d \rho \right) ^{-1}_{\rho=0}
= \theta_h/\alpha.
\label{size}
\end{equation}
The solution of the boundary value problem
(\ref{eq}), (\ref{Boundary}) can be 
readily found from a set of
solutions of the Cauchy problem $\theta (\rho; \alpha)$.
Namely, the solution is given by the profile $\theta (\rho)$
which crosses line $\rho = R_d$
at the angle $\beta = \arctan (g(\theta, R_d))$ (Fig. \ref{shoot1} (b)).
The family of the solutions $\theta (\rho; \alpha)$
of the Cauchy problem 
establishes the connections between
the equilibrium vortex size 
($R_0 \propto 1/\alpha$), 
the disk radius $R_d$, 
and the anchoring energy $g(\theta, R_d)$.
The trajectories $\theta(\rho; \alpha)$ 
in Fig.~\ref{shoot1} (b) visually demonstrate 
how the vortex core size depends 
on the disk radius $R_d$ and the boundary anchoring 
at the edges.
This dependence should be noticeable
in \textit{small} disks with radii
comparable to the exchange length $l_e$.
In real magnetic nanodisks 
one usually has $R_d \gg l_c$.\cite{Shinjo00}
In this case
the solutions of Eqs.~(\ref{eq}), (\ref{Boundary})
are very close to separatrix lines
and the equilibrium core 
retains a fixed shape and size
almost identical to the separatrix solution
$\tilde{\alpha}$ and $ \theta_h$,
largely independent from the radius $R_d$ 
and the anchoring energy.
In this connection see an interesting
discussion on vortex core sizes
in Ref.~\cite{Li06}.

The independence of the core structure
on boundary conditions 
allows to neglect effects imposed 
by the anchoring energy at the edges. 
Thus, we consider here the problem
with \textit{free} boundary conditions,
$g(\theta,R_d) =0$.
Eqs. (\ref{eq}), (\ref{Boundary}) include
two independent material parameters,
$D/D_0$ and $H/H_a$.
For fixed values of these control
parameters the solutions of the
vortex profile  
have been derived by using the  following 
numerical procedure.
The Cauchy problem (\ref{eq}), (\ref{Cauchy}) was
solved by the Runge-Kutta method. 
Through repeated calculations for varying values $\alpha$, 
the correct trajectory was searched by
'shooting' at the boundary condition value (\ref{Boundary}).
After that the profiles have been
improved by a relaxation calculation 
using a finite-difference method
for the boundary value problem, for details see Ref.\cite{JMMM94}.

The chirality of a non-collinear structure
can be measured from the strength of the twist or helical
rotation of the magnetization,
$\mathbf{m}\cdot(\nabla\times\mathbf{m})$. 
For the radial vortex structure, 
the local twist is given by the expression
\begin{equation}
\label{Localchi}
\tau=\left(\frac{d \theta}{d \rho}+\frac{1}{\rho}\cos\theta\sin\theta\right)\,.
\end{equation}
The sign of this expression measures
the local and helical chirality in the structure.
The comparison with Eq. (\ref{density2}) shows that the local
twist is equivalent to the local density of the DM energy.
In particular, for $0\le \theta \leq \pi/2$ and $\theta_{\rho}>0$ 
the local chirality of the helical structure is positive.
Alternatively, the local chirality can be measured
by the $\mu=z$ component of the chiral current 
$j_{\mu}=(1/(8\pi^2)\epsilon_{\mu\nu\lambda}\mathbf{m}\cdot(\partial_{\nu}\mathbf{m}\times\partial_{\lambda}\mathbf{m})$. 
The evaluation for the vortices structure 
gives a chirality
\begin{equation}
\chi=(\theta_{\rho}/(2\pi\rho))\sin\theta\,. 
\end{equation}
Both the sign of $\tau$ and $\chi$ can be used to determine
the local handedness in the vortex structure.
In particular, a change of slope for the profile from 
$\theta_{\rho}>0$ to $\theta_{\rho}<0$
changes the chirality of the magnetization structure (Fig. \ref{profilev}, Inset).

\section{ Results}

In this section we present the results 
for the vortex core structures first from exact
numerical calculations. 
Then an analytical ansatz is discussed which 
offers qualitative insight on the dependence
and mechanism by which DM couplings influence
the core structure of vortices.
Finally the stability and static distortion modes
of the vortex solutions are investigated. 

\subsection{Vortex solutions and magnetization profiles}

\begin{figure}
\includegraphics[width=7.5cm]{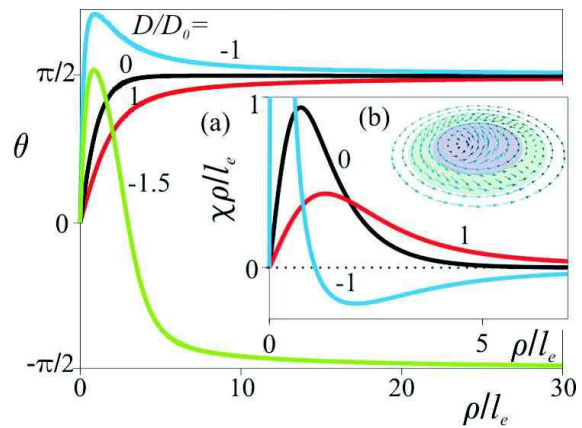}
\caption{
Typical solutions $\theta (\rho)$ for different
values of $D$ in zero applied field.
The vortices widen when their chirality $c=1$
coincides with the chirality of the DM
interaction, $D>0$. 
In the opposite case, $D<0$, the
vortex cores shrink, and the 
local chirality $\chi(\rho)$ changes
its sign outside the vortex core (Inset a).
The core with unfourable chirality 
is encircled by a ring with negative local 
polarization and favourable chirality (Inset b).
For strong negative DM coupling ($D/D_0=-1.5$) the 
circulation of the projected magnetization in the plane
changes the sense of rotation from positive in the core to 
negative at the edge, $\theta \rightarrow -\pi/2$ at $R_d$.
\label{profilev}
}
\end{figure}

\begin{figure}
\includegraphics[width=7.5cm]{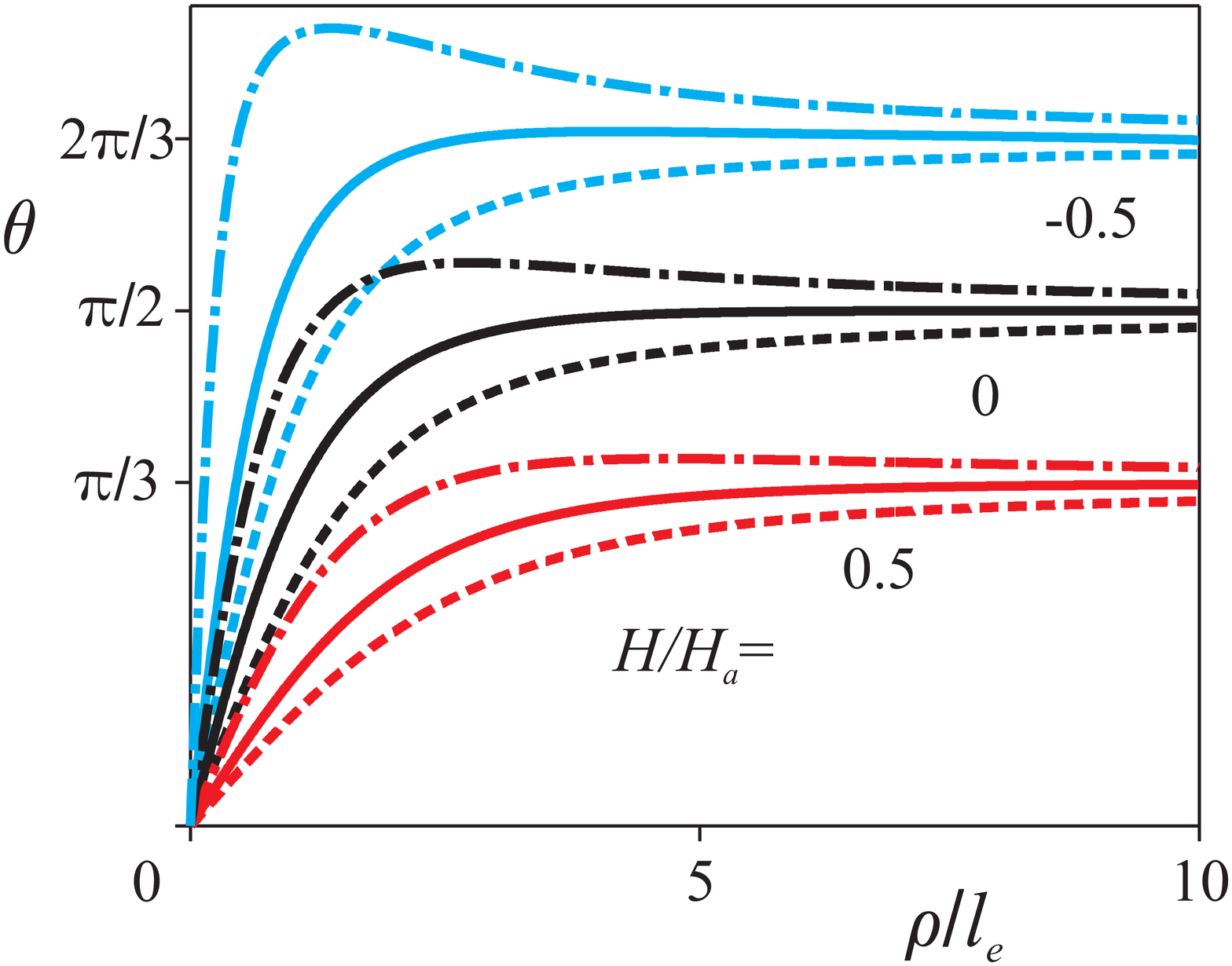}
\caption{
Typical solutions $\theta(\rho)$ for different
values of the applied field and the reduced
values of the Dzyaloshinskii constant:
$D/D_0$ = 0 (solid lines), 0.5 (dashed lines),
-0.5 (dash-dotted lines).
\label{profH}
}
\end{figure}

Typical solutions of Eq. (\ref{eq}) for different 
values of $D$  in zero field
are shown in Fig.~\ref{profilev}.
The effect of applied perpendicular fields
on the core structure is demonstrated 
in Fig.~\ref{profH}.
All reported results from the numerical solution
of the boundary problem (\ref{eq}),(\ref{Boundary})
are for a fixed disk radius $R_d = 30 l_e$.
As discussed, the solutions well represent
the vortex core properties for any $R_d \gg l_e$.
Vortex profiles
$\theta(\rho)$ have an arrow-like shape.
For $D > 0$
the angles $\theta$ vary monotonically
from zero at the vortex axis to 
$\theta_h$ (\ref{thetah}).
In this case the magnetization
has everywhere a local chirality favoured by the Dzyaloshinskii-Moriya
interactions.
The core size widens with increasing $D$.

\begin{figure}
\includegraphics[width=7.5cm]{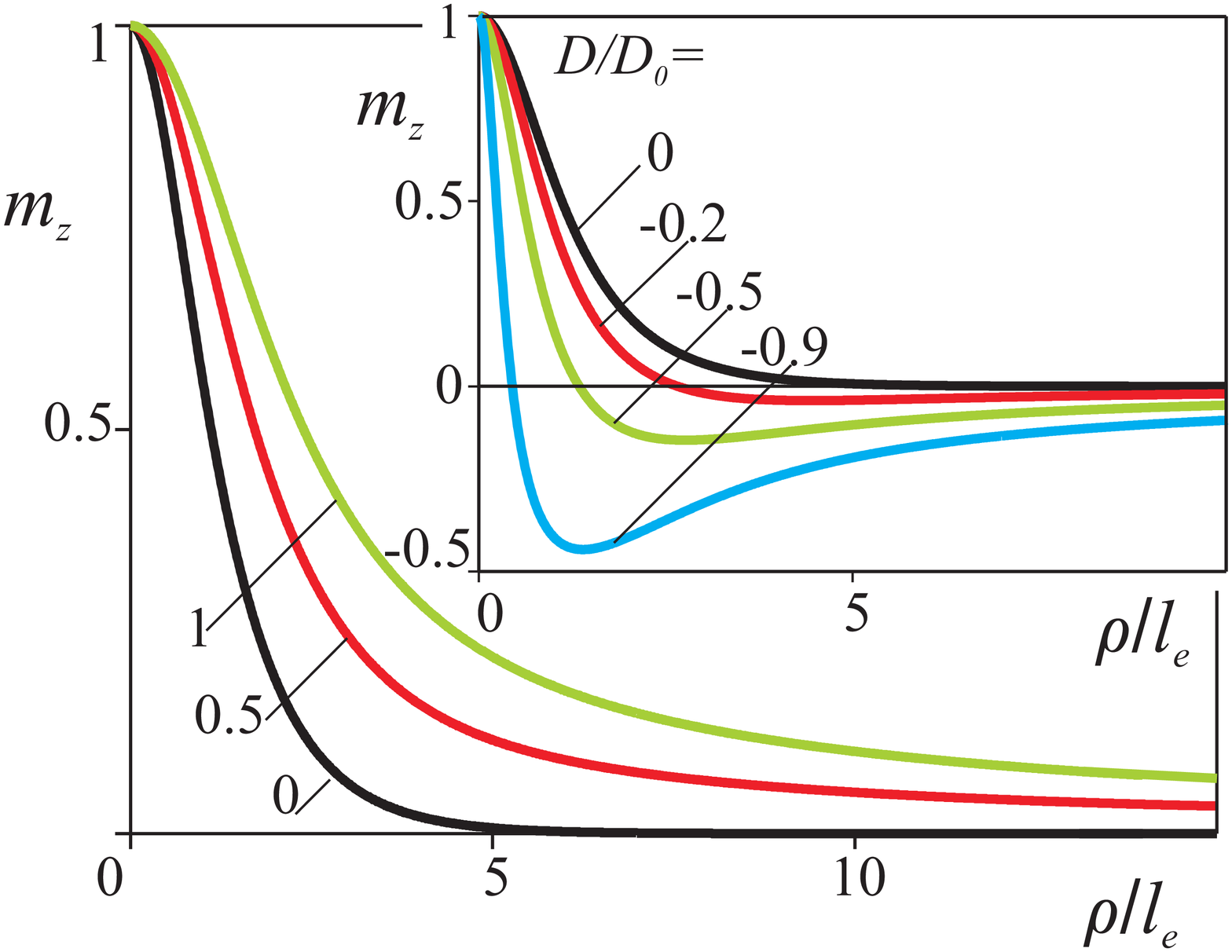}
\caption{
Magnetization profiles $m_z(\rho/l_c)$
for different values of $D$.
For $D > 0$ 
the perpendicular magnetized central
spot broadens with increasing D.
For negative $D$
an increasing $|D|$ widens
the ring with negative $m_z$
and compresses a central spot
with positive $m_z$ (Inset).
\label{magn}
}
\end{figure}
\begin{figure}
\includegraphics[width=7.5cm]{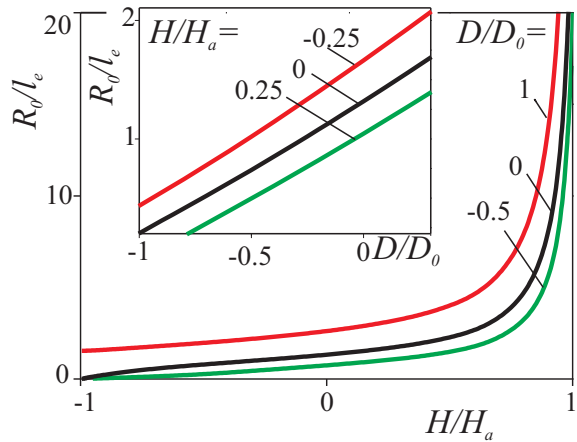}
\caption{
The equilibrium sizes of vortex core
$R_0$ (\ref{size}) as functions 
of reduced magnetic field $H/H_a$
for different values of $D/D_0$.
Inset shows $R_0$ as
functions of $D/D_0$
for different values of 
the applied magnetic fields.
\label{sizes1}
}
\end{figure} 

For negative $D$ 
the magnetization in the core has unfavourable
chirality. 
As a result in a certain point
$\rho_r$ with $\theta (\rho_r) > \theta_h$ the
profile $\theta(\rho)$ goes through a maximum, $\theta>\pi/2$ in zero field,
the slope $\theta_{\rho}$ becomes negative, and the
magnetization structure changes its local chirality.
After that, in the range
$ \rho_r < \rho < R_d$ the polar angle
$\theta (\rho)$ monotonically approaches
the limiting value $\theta_h$.
As the magnitude of the negative $D$ increases 
these vortices transform into those
with "reverse" rotation to $-\theta_h$
(profile with $ D = - 1.5$ in Fig. \ref{profilev}).
For $D < 0$ the vortex core consists of 
a narrow internal part ($\rho < \rho_r$)
and the adjacent ring with a reverse magnetization
rotation. 

Peculiarities of the vortex profiles for
different sign of $D$ are reflected in
their magnetization distribution
(Fig.~\ref{magn}).
Positive $D$ increases the width of 
the central spot with $m_z > 0$ 
(Fig.~\ref{magn}). 
As a result
the total perpendicular magnetization
of the disk
$ <m_z > = \int_0^{R_d} m_z(\rho) \rho d \rho$ 
is larger in vortices of the right positive chirality.
Negative $D$ squeezes the central magnetization
core with $m_z >0$ and widens the adjacent ring
with negative perpendicular magnetization 
($m_z < 0$) (Fig.~\ref{magn}, Inset).
The overall behavior of the vortex core sizes
is summarized in Fig.~\ref{sizes1} 
by displaying the dependence of the core radii $R_0$,
as defined in Eq.~(\ref{size}), on the external
field for different zero, positive, and negative
values of $D$.
This dependence is weak and 
almost linear up to large fields $H/H_a \rightarrow 1$,
where the transition into the homogeneously magnetized
state takes place.
The Inset of Fig.~\ref{sizes1} shows 
that the dependence of $R_0$ on $D$ is almost linear
both for zero field and for not too large positive and
negative fields.

\subsection{Linear ansatz and analytical results for a vortex core}

Vortex profiles with strongly localized
arrow-like cores (Fig.~\ref{profilev}) 
can be described
by a linear ansatz for the core and a flat part, 
\begin{eqnarray}
&& \theta = \theta_h (\rho/R), \quad  0 < \rho < R,
\nonumber \\
&& \theta = \theta_h, \quad  R < \rho < R_d.
\label{ansatz}
\end{eqnarray}
Integration of the energy functional Eq. (\ref{density2}) 
with the ansatz $\theta (\rho)$ (\ref{ansatz})
leads to the following expression for 
the vortex energy as a function of $R$,
\begin{eqnarray}
\mathcal{E} (R) = \mathcal{K} R^2 - \mathcal{A} \ln R
- \tilde{c}\mathcal{D} R \,,
\label{ansatzE}
\end{eqnarray}
where $\mathcal{K} = \pi  g_0 (H)K$,
$\mathcal{D} = 4 \pi g_1 (H) D$,
$\mathcal{A} = 2\pi g_2 (H)A$,
and
$g_0(H) = ( \theta_h^2(1+2\cos^2 \theta_h)
-3 \theta_h \sin 2\theta_h -7 \cos^2 \theta_h +8 \cos \theta_h-1)/(2 \theta_h ^2)$,
$g_1(H) = (\sin^2 \theta_h -\theta_h  \sin 2 \theta_h+\theta^2_h)/(4\theta_h)$,
$g_2(H) =  \sin^2 \theta_h $,
and $\theta_h$ is the solution of Eq. (\ref{thetah}). 
In Eq. (\ref{ansatzE}) we omit terms independent on parameter $R$.
Minimization of energy (\ref{ansatzE}) yields the equilibrium
radii of the core for positive and negative $D$
\begin{eqnarray}
R_{1,2} =  u(H) l_e
\left[ \sqrt{\left(\frac{D}{D_0} \right)^2 + v^2(H)} 
+ \tilde{c} \frac{|D|}{D_0} \right],
\label{ansatzS}
\end{eqnarray}
where $u(H)=g_1(H)/g_0(H)$, $v(H)=\sqrt{g_0(H)g_2(H)}/ g_1(H)$.
Particularly, at zero field
$u(0)=1.856, v(0)=0.988$.

Eqs. (\ref{ansatzE}), (\ref{ansatzS}) offer an important
insight into the physical mechanism that underlies
the  formation of the vortex states.
The exchange energy of the vortex core does not depend
on its size.
This reflects a general property of vortex
and 2D Skyrmionic states.\cite{Derrick64,Nature06}
The exchange energy of the adjacent ring ($\propto - A \ln R$)
favours the extension of the vortex cores while
the magnetic anisotropy energy $\propto  K R^2$ tends to
compress  them. 
For $D =0$  the balance between these energy contributions
yields the equilibrium core sizes 
$ R \propto \sqrt{A/K}$, equal for the vortices of
opposite chirality.
Finite values of $D$ violate chiral symmetry
of the solutions (\ref{ansatzS}) stabilizing
vortices with different sizes of the core.
The difference between core sizes in
vortices with different chirality
can be readily derived from 
Eq. (\ref{ansatzS}),
\begin{eqnarray}
\Delta R = |R_{1}-R_{2}| = 2 l_e
 \frac{|D|}{D_0} u(H) \,.
\label{ansatzS2}
\end{eqnarray}
The numerical calculations of $\Delta R$ also reveal
a linear relation  $\Delta R = a (D/D_0)$,
where the coefficient $a$ depends on 
the applied field.
Particularly  $a(H/Ha)$ has the following
values:  $a(0) = 2.4$, $a(0.25) = 2.66$,
and $a(-0.25) = 2.24$.
%


\subsection{Radial stability of the solutions}
To study radial stability of vortices
we consider a small arbitrary radial distortion $\xi(\rho)$ 
of the solutions $\theta(\rho)$ of Eqs. (\ref{eq}),
(\ref{Boundary}) with the boundary conditions
 $\xi(0)=\xi(R_d)=0$.
We insert $\tilde{\theta}(\rho)=\theta(\rho)+\xi(\rho)$ 
into energy functional Eq. (\ref{density2}) and keep only terms 
up to second order in $\xi(\rho)$.
Because $\theta(\rho)$
is the solution of the boundary value problem 
the first-order term must 
vanish, and one obtains  
$E=E^{(0)}+E^{(2)}$
where $E^{(0)}$ is the equilibrium energy, and
\begin{eqnarray}
E^{(2)}=2\pi\int^{R}_{0}\left[A \left(\frac{d \xi}{d \rho}\right)^2+G(\rho)\xi^2\right]\rho\, d \rho
\label{Dist}
\end{eqnarray}
with
\begin{eqnarray}
\label{G}
G(\rho)&=&\frac{A}{\rho^2}\cos2\theta-K\cos2\theta \\
& & +\frac{1}{2}HM_s\cos\theta+\frac{D}{\rho}\sin2\theta \,.\nonumber
\end{eqnarray}
\begin{figure}
\includegraphics[width=7.5cm]{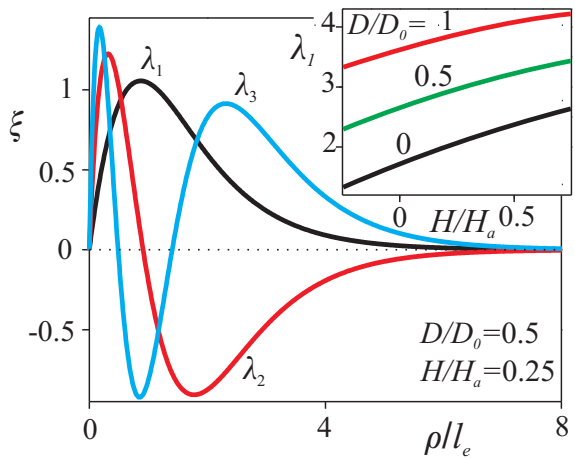}
\caption{
The first three excitation modes for $D/D_0=0.5$ and $H/H_a=0.25$ and
corresponding eigenvalues: $\lambda_1=2.966$, $\lambda_2=8.428$, $\lambda_3=15.685$.
The structure is radially stable because the smallest eigenvalue $\lambda_1$ is positive.
Inset shows the first eigenvalue as a function of the applied field and for different values of 
$D/D_0$.
\label{stab}
}
\end{figure}

\begin{figure}
\includegraphics[width=7.5cm]{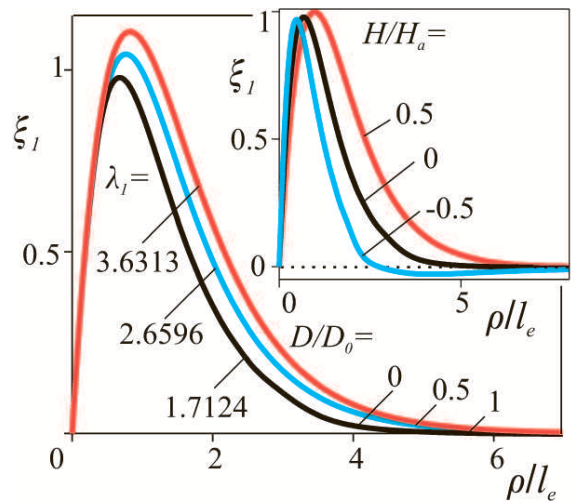}
\caption{
The first excitation mode $\xi_1 (\rho)$ 
at zero field and for different
values of the Dzyaloshinskii constant.
Inset shows $\xi_1 (\rho)$  for $D = 0$
and different values of the applied field.
Corresponding eigenvalues $\lambda_1(H/H_a)$:
$\lambda_1(-0.5) = 0.9556$, $\lambda_1(0) = 1.7124$, $\lambda_1(0.5) = 2.3644$.
\label{stab1}
}
\end{figure}
The stability problem is reduced to the solution of 
the spectral problem for functional (\ref{Dist})
(for details see Ref. \cite{JMMM94}).
By expanding $\xi(\rho)$ in a Fourier series
\begin{eqnarray}
\xi(\rho)=\sum^{\infty}_{k=1}b_k\sin(ak\theta(\rho)),
\label{Fourier}
\end{eqnarray}
where $a=\pi/\theta_h$
the perturbation energy $E^{(2)}$(\ref{Dist}) 
can be reduced to the following quadratic form
\begin{eqnarray}
E^{(2)}=\sum^{\infty}_{k,j=1}A_{kj}b_kb_j
\label{Quadr}
\end{eqnarray}
with
\begin{eqnarray}
A_{kj}=\int^{R}_{0}[A a^2kj\cos(ak\theta)\cos(aj\theta)+\\ 
+G(\rho) \sin(ak\theta) \sin(aj\theta)] \rho d \rho. \nonumber
\label{Matrix}
\end{eqnarray}
Radial stability of the solution is determined
by the sign of the smallest eigenvalue 
$\lambda_1$ of matrix \textbf{A}:
if   $\lambda_1 > 0$ the solution $\theta(\rho)$ is stable
with respect to  perturbations $\xi (\rho)$,
and the solutions are radially unstable if $\lambda_1$
is negative.

Numerical calculations 
demonstrate radial stability
of vortex solutions for positive $D$ in the whole range
of the magnetic fields where these solutions exist.
For $H =0.25$, $D =0.5$ the first three excitation modes
$\xi (\rho)$ and their eigenvalues are shown in
Fig. \ref{stab}.
The variation of the first eigenfunctions and eigenvalues 
under the influence of the applied field
and $D > 0$ is shown in Fig. \ref{stab1}.
The lowest perturbation eigenfunctions
are connected with an expansion or compression
of the vortices.
The eigenvalues and the eigenfunctions 
correspond to certain magnetic resonance modes 
associated with radial deformations of the vortex core.

For $D < 0$ radial stable solutions exist
only below certain critical strength of 
the Dzyaloshinskii constant $ |D| < |D_{cr}|$.
For $ |D| > |D_{cr}|$ vortex solutions
are radially unstable.
Detailed investigations of vortex
structures for $D < 0$ beyond this threshold 
and their stability is beyond this 
contribution restricted to axisymmetric
simple vortex structures.

\section{Conclusions}

We have investigated the influence of 
induced Dzyaloshinskii-Moriya
interactions on the equilibrium vortex 
parameters in magnetic nanodisks.
Both numerical and analytical calculations demonstrate
strong dependencies of the vortex structures (Fig. \ref{profilev}), 
magnetization profiles (Fig.~\ref{magn}),
and core sizes (Fig. \ref{sizes1})
on the strength and sign of the Dzyaloshinskii-Moriya coupling.
Thus, by switching the chirality of a vortex 
a change of the vortex profile, core size, and the perpendicular
magnetization takes place in the presence of a surface-induced 
Dzyaloshinskii-Moriya coupling.
Existing experimental imaging techniques should already be 
sufficient to investigate these differences in vortex states 
with different chirality.\cite{Raabe00,Schneider00,Schneider01,Curcic08,Shinjo00,%
Wachowiak02,Li06,Tanase09,Weigand09,Loubens08}
The  calculated relations between 
strength of the Dzyaloshinskii-Moriya
interactions and vortex core sizes 
(Fig.~\ref{sizes1} and Eqs. (\ref{ansatzS}), (\ref{ansatzS2}))
provide a method for experimental determination of 
the Dzyaloshinskii constant $D$.

From the theoretical side our results obtained within
a simplified micromagnetic model can be extended.
The calculated dependence of the excitation modes 
in Fig.~\ref{stab} on the Dzyaloshinskii-Moriya interactions
indicate that magnetic resonances of vortex cores, or more
generally dynamical effects, may provide a route to investigate
chiral symmetry breaking in film elements with vortex states.
Many recent experimental \cite{Tanase09,Weigand09,Yamada07} 
and theoretical \cite{Komineas07,Gaididei08,Liu09} studies
have been performed on vortex-core dynamics and magnetization
reversal in nanodisks  under 
the influence of magnetic fields or current pulses.
Effects of surface-induced Dzyaloshinskii-Moriya 
interactions have not been considered yet 
for these nanomagnetic processes. 
Our investigation on static vortex structures
is a first step towards detection of these chiral
couplings in magnetic nanodisks.
\begin{acknowledgments}
The authors are grateful to S. Bl\"ugel, N. S. Kiselev and S. Komineas
for fruitful discussions.
A.N.B.\ thanks H.\ Eschrig for support and
hospitality at IFW Dresden. 
\end{acknowledgments}

\end{document}